# Phenomenological model of high-$T_c$ superconductivity: a MCS model

A. Mourachkine

*Université Libre de Bruxelles, Service de Physique des Solides, CP233, Boulevard du Triomphe, B-1050 Brussels, Belgium*

Since the introduction of a MCS (Magnetic Coupling between Stripes) phenomenological model [cond-mat/9902355, to be published in J. Superconductivity] for hole-doped cuprates many new experimental data have been presented in the literature as evidence in support of the MCS model. We consider here recent data and the MCS model which is based upon experimental facts, namely, the presence of (i) stripes; (ii) spin fluctuations, and (iii) two order parameters (for pairing and for long-phase coherence) in hole-doped cuprates. We discuss also the superconductivity in the s-wave $Nd_{2-x}Ce_xCuO_4$ cuprate.

## I. INTRODUCTION

Since the discovery of high-temperature superconductors (HTSC) in 1986 by Bednorz and Müller[1] there is no consensus on the mechanism of superconductivity (SC) in copper-oxide materials. There is a clear microscopic difference between conventional SCs and HTSC, namely that they have different origins and that different criteria are required for the HTSC than for the classical SCs. The normal-state properties of SC cuprates are also unique, such as the presence of pseudogap, the linear dependence of in-plane resistivity on temperature *etc*. It is widely believed that the $CuO_2$ planes are responsible for the occurrence of SC in cuprates. The parent compounds of the HTSC are antiferromagnetic (AF) insulators with a large value of the exchange constant. Many experiments show some universality of physical properties across widely different cuprates with different values of $T_c$. As for the bulk characteristics, the controlling factor in cuprates seems only to be the hole density, $p$, in $CuO_2$ planes.[2] The definitive confirmation of the existence of the $d_{x^2-y^2}$ (hereafter, d-wave) order parameter (OP) in hole-doped cuprates is a great advance in recent years.[3] However, the interpretation of some experimental data suggests the presence of s-wave component in hole-doped cuprates.[4] At the same time, there are some differences among different families of cuprates, the most distinctive one is the s-wave symmetry of total OP in $Nd_{2-x}Ce_xCuO_4$ (NCCO) which is widely believed to be an electron-doped cuprate.[5]

There is clear evidence for charge stripe[6] formation in $La_{1.6-x}Nd_{0.4}Sr_xCuO_4$ (Ref. 7) and $YBa_2Cu_3O_{6+x}$ (YBCO).[2,8,9] There are also strong indications that stripes exist in $Bi_2Sr_2CaCu_2O_{8+x}$ (Bi2212).[10] Holes induced in $CuO_2$ planes segregate into periodically-spaced stripes that separate AF isolating domains. By using this experimental fact Emery, Kivelson and Zachar presented a stripe model of underdoped HTSC.[11] The main feature of their model is a spinon SC which occurs on one-dimensional charge stripes in $CuO_2$ planes at $T_{pair} > T_c$. Spinons are neutral fermions, and they are not impended by the Coulomb interactions. Spinons form pairs on stripes due to pair hopping between the stripe and its AF environment.[11] In the stripe model, the coherent state of the spinon SC is established at $T_c$ by the Josephson coupling between charge stripes carrying the spinon SC. However, the stripe model can not explain the presence of two distinct energy scales in hole-doped cuprates,[12,13] which are both SC-like, and one of them has magnetic origin.[14] Taking into account the latter fact we found that the stripe model may fit the data after a serious modification, namely, if instead of the Josephson coupling between stripes we introduce a magnetic interstripe coupling due to spin fluctuations (spin-waves).[15] We call this phenomenological model as a MCS (Magnetic Coupling between Stripes). Thus, the MCS model is entirely based upon experimental facts. Moreover, we found that many other experimental data can be explained by using the MCS model.[16] In addition to this, the "modified" MCS model may explain the SC in s-wave NCCO.[15,16] However, we understand that, probably, there is another model of the HTSC which can explain experimental data better. At this moment, we are not aware of such model.

In the MCS model, the coherent state of the spinon SC is established due to spin-waves which are excited into local AF domains by dynamically fluctuating stripes. Thus, in the MCS model, the SC in cuprates has two mechanisms: along charge stripes and perpendicular to stripes. Charge carriers exhibit different properties in different directions: *fermionic* along charge stripes and *polaronic* perpendicular to stripes. According to the MCS model, the Anderson's theorem for cuprates is different: *any impurity (magnetic or nonmagnetic) doped into $CuO_2$ planes affects both the $T_c$ value and density of states (DOS)*. In this paper, we present the MCS phenomenological model of the HTSC and recent experimental data obtained by different techniques as an experimental evidence in support of the MCS model.

The structure of the paper is as follows. Stripes in cuprates are discussed in Section II. In Section III, we consider spin fluctuations in hole-doped cuprates. The order parameters for pairing and long-range phase coherence in hole-doped cuprates are discussed in Section IV. The MCS phenomenological model of the SC in hole-doped cuprates is described in Section V. In Section VI, we present some experimental data obtained on hole-doped cuprates by different techniques, which can be explained by using the MCS model. In Section VII, we focus attention on the SC in the s-wave NCCO cuprate. The Anderson's theorem for cuprates is considered in Section VIII. The final discussion is presented in Section IX. The conclusions can be found in Section X.

**II. STRIPES**

Static charge stripes[6] are found in nickelates[17] and manganates.[18] The dopant-induced holes choose to segregate in these materials into periodically-spaced stripes which separate AF insulating domains. The clearest evidence for stripe correlations in cuprates has been provided by neutron and x-ray scattering studies of Nd-doped LSCO.[7] Study of the charge modulation has been become possible with the discovery that, for hole concentration of 1/8 per Cu atom, stripes can be pinned by the structural modulation induced by partial substitution of the ion $Nd^{3+}$ for $La^{3+}$.[7] Recently, many experiments have been also provided evidence ($^{63}$Cu nuclear quadrupole resonance (NQR),[19] EXAFS,[20] x-ray,[21,22] muon-spin-rotation (μSR)[23,24] and transport[22,25] measurements) for stripe correlations in pure or doped LSCO. Very recently, through neutron and x-ray diffraction and transport measurements on Nd-doped LSCO, it is showed that the charge stripes which order within the $CuO_2$ planes at low temperature are intrinsically metallic, contrary to what one would expect for conventional charge-density-wave (CDW) order.[22] From these measurements, it was concluded that, for other SC cuprates, the lack of an upturn in the resistivity at low temperature does not necessarily indicate the lack of charge stripes, but rather the absence of stripe pinning.[22] Experimentally, it is observed that stripes in cuprates are charge-driven, *i.e.* the charge-peaks appear at higher temperature than the magnetic peaks.[26] Stripes in cuprates are "vertical" (along crystal axes) while they are "diagonal" (at 45° to an axis) in the nickelates.[26] From $^{63}$Cu NQR measurements, the stripes in cuprates are not static but slowly fluctuating with the characteristic time of 10-12 μsec.[19] The dynamical nature of stripes in cuprates helps them to avoid a CDW instability.[11]

Recently, there are many different kinds of evidence for stripe correlation in YBCO. Phonon measurements on YBCO reveal the presence of charge fluctuations which serve as a direct evidence for the existence of a dynamic striped phase in YBCO.[9] The inferred periodicity of charge fluctuations is that expected if the charge is located in the domain walls separating the spin stripes. The stripe formation and spin fluctuations in local AF domains, which we will consider in the next Section, relate to each other.[8] In YBCO, the first evidence for stripe correlations came from spin dynamics revealed by neutron scattering studies.[2,8,27,28] The found value of the incommensurability, ~ 1/8, is very close to the value expected from the hole concentration.[7,26,28] These results demonstrate that the spin dynamics do not depend on the details of Fermi surface but have an analogous form to that for the stripe structure in LSCO. Thus, they provide a very direct connection between the spin correlations in SC YBCO and LSCO. Recent magnetoresistance measurements on *undoped* YBCO show clearly the presence of charge stripes which can be directed with an external magnetic field.[29]

There is less direct evidence for stripe correlations in other cuprates than in LSCO and YBCO. In transport and μSR measurements, the anomalous suppression of SC has been found also in Zn-substituted Bi2212 around $p$ ~ 1/8 per Cu.[10,30] Thus, this "1/8 effect" is now recognized to be common to cuprates and is a direct evidence for the presence of charge stripes in LSCO, YBCO and Bi2212. It seems that stripes are intrinsic in all cuprates, however, there is no consensus how stripes and the origin of the HTSC relate to each other.[31]

### III. SPIN DYNAMICS IN HOLE-DOPED CUPRATES

Neutron-scattering, nuclear magnetic resonance (NMR) and μSR studies showed that the local AF spin correlations coexist with the SC in the layered cuprates.[7,24,26] Charge segregation provides a way to reconcile these properties. Fluctuations of the Cu spins in AF domains give rise to magnetic excitations, and might mediate the electron pairing that leads to SC. Indeed, recent neutron studies reveal that the magnetic fluctuations at low frequencies are beginning to achieve the same universality across widely different materials with different values of $T_c$.[2,8,28,32,33] As for the bulk characteristics, the controlling factor in cuprates seems only to be the hole density in $CuO_2$ planes.[2]

So-called magnetic "resonance peak" found earlier in YBCO[34-37] by inelastic neutron scattering (INS) has been recently detected in Bi2212.[38] The discovery of the resonance peak in Bi2212 points out that the resonance peak is an *intrinsic* feature of the SC in the double-layer *hole*-doped cuprates studied so far. The resonance-peak position, $E_r$, increases first when the doping increases and then decreases again in the overdoped regime.[14,36] The temperature dependence of the resonance peak demonstrates that this peak is intimately related to the establishment of SC.[37] It is important to note that the magnitude of $E_r$ is in quantitative agreement with the condensation energy of YBCO.[39,40] The resonance peak has been also observed by INS in a heavy fermion compound[41] $UPd_2Al_3$ for which spin fluctuations are believed to mediate the pairing interactions.[42,43] It is also important to note that the SC in $UPd_2Al_3$ coexists with the AF order like in the cuprates.[41,42] By making a parallel between the cuprates and heavy fermion compound $UPd_2Al_3$, the latter suggests that spin fluctuations mediate the electron pairing that leads to SC in hole-doped cuprates. The SC mediated by spin fluctuations implies that the OP has the d-wave symmetry.[41-45]

Some theoretical calculations show that spin fluctuations are too weak to produce the HTSC pairing mechanism. However, the comparison of INS and infrared spectroscopy data shows that (i) the conducting carriers are strongly coupled to a resonance peak, and (ii) the coupling strength inferred from those results is sufficient to account for high value of $T_c$.[46] Thus, recent INS studies show that AF spin exchange interactions seem to be intrinsic in hole-doped cuprates[2,8,32,33] and, most likely, responsible for the high-$T_c$ SC mechanism.[14,33,44-46]

## IV. ORDER PARAMETERS FOR PAIRING AND COHERENCE

In this Section, we discuss the OP for pairing and OP for long-range phase coherence in hole-doped cuprates, their symmetries, origins and temperature dependencies.

The SC in metals requires the presence of the Cooper pairs and the phase coherence among them.[47] In the BCS theory for conventional SCs,[47] the pairing mechanism and mechanism of establishment of the long-range phase coherence are identical: two electrons in each Cooper pair are attracted by phonons, and the phase coherence among the Cooper pairs is established also by phonons. Both the phenomena occur almost simultaneously at $T_c$. In SC copper-oxides, there is a consensus that, in the underdoped regime, the pairing occurs above $T_c$ when the long-range phase coherence is established, $T_{pair} > T_c$.[11-16,48-50] Moreover, there are strong indications that the origins of the two mechanisms are different.[13-16] Firstly, the magnitudes of the OPs for pairing and long-range phase coherence in hole-doped cuprates have different dependencies on hole concentration in $CuO_2$ planes.[11-16] The magnitude of the OP for long-range phase coherence, $\Delta_c$, which is proportional to $T_c$, has the parabolic dependence on $p$,[12,13,51] while the magnitude of the OP for pairing, $\Delta_p$, increases linearly with the decrease of hole concentration.[12,13,52] Figure 1 shows the phase diagram for the two OPs in hole-doped cuprates, which is *common* for hole-doped cuprates.[12,13] In Fig. 1, the $\Delta_c$ scales with $T_c$ as $2\Delta_c/k_B T_c = 5.45$.[13] It is worth noting that both the two gaps (the magnitudes of the OPs) are SC-like.[14] Secondly, an applied magnetic field affects the two OPs in cuprates differently: the long-range phase coherence disappears first, while in order to break the Cooper pairs the magnitude of magnetic field must be much higher.[16] However, in spite of the facts which show that the origins of the $\Delta_c$ and $\Delta_p$ are different, *there is no* consensus on the origins of these two OPs.

In our previous studies, we considered in details the symmetries[53] of the two OPs and its origins.[54,14] We found that, most likely, the OP for long-range phase coherence has the d-wave symmetry while the OP for pairing has an anisotropic s-wave symmetry.[53,54] The origin of the coherent OP is magnetic due to spin fluctuations into local AF domains (see Section III).[14,54] The pairing OP is responsible for coupling of spinons.[54,55] In our previous works, we discussed also the temperature dependencies of the $\Delta_c$ and $\Delta_p$. We showed that, below $T_c$, the temperature dependence of the $\Delta_p$ is temperature independent and lies above the BCS[47] temperature dependence.[12,16,56] There is no changes in the magnitude of the $\Delta_p$ at $T_c$. On the other hand, the temperature dependence of the $\Delta_c$ lies below the BCS[47] temperature dependence.[12,16,56] It is important to note that the magnitude of the OP for long-range phase coherence becomes zero at $T_c$ with increase of temperature in terms of long-range scale, however, since the magnetic resonance peak heavily damped has been observed above[32] $T_c$ there are *local* fluctuations of the coherent OP between $T_c$ and $T_{pair}$. The magnitude of the $\Delta_p$ becomes zero at $T_{pair}$.

Let us consider briefly the shapes of the total OP, $\Delta_c$ and $\Delta_p$ in Bi2212, which have been discussed in details elsewhere.[53,54] The magnitude of total OP is equal, in general, to $\Delta = (\Delta_c^2 + \Delta_p^2)^{1/2}$.[57] Figure 2(a) shows tunneling data obtained on

Bi2212 single crystals with $T_c$ = 85 K ($p/p_m$ = 1.2) by angle-resolved tunneling measurements.[58] The tunneling gap shown in Fig. 2(a) corresponds to the magnitude of the total OP in Bi2212. Figure 2(b) shows separately the shapes of the $\Delta_c$ and $\Delta_p$ in Bi2212.[53,54] In Fig. 2(b), the pairing OP in Bi2212 has an *anisotropic* s-wave symmetry. Such combination of the two OPs in Bi2212 is supported by tunneling measurements[59] on Pb-doped Bi2212 which clearly demonstrate that a tunneling gap with the maximum magnitude doesn't relate to $T_c$ (*i.e.* to $\Delta_c$). In the Pb-doped Bi2212 cuprate in which $CuO_2$ planes are, in the first approximation, unaffected by Pb-doping since Pb atoms reside between $CuO_2$ bilayers, the maximum of the gap magnitude which is located in Cu-Cu direction [see Fig. 2(a)] remains unchanged in comparison with pure Bi2212 case, while the minimum of the gap magnitude observed along Cu-O-Cu bond direction decreases proportionally to the decrease in $T_c$.[59] Pb atoms doped in Bi2212 affect only the OP for long-range phase coherence keeping the pairing OP unchanged. This implies that the pairing OP is present in Cu-Cu direction and the coherent OP is dominant in Cu-O-Cu bond direction. This is in a good agreement with Fig. 2(b).

The two OPs has been experimentally observed in tunneling measurements on pure and Ni-doped Bi2212.[12] Independently from these measurements, the analysis of Andreev reflection, penetration depth, Raman scattering, angle-resolved photo-emission and tunneling measurements made by Deutscher showed the presence of two distinct energy scales in hole-doped cuprates.[13] Surprisingly, the two phase diagram obtained independently from each other practically coincide.[12,13,60] The tunneling measurements showed also the presence of a sub-gap in Bi2212, which was attributed to a g-wave SC gap.[12] Since theoretically, the d- and g-wave magnetically mediated SCs may co-exist[44] *it is possible* that, in hole-doped cuprates, there exists the g-wave OP[12,15,53] with the maximum magnitude of 30%-50% from the maximum magnitude of the d-wave component.[44] The g-wave component does not change the maximum magnitudes of the two OPs shown in Fig. 2(b), it just adds some "weight" between the maxima of the two OPs.[53]

In summary, the analysis made in our previous studies, shows that, most likely, the Cooper pairs in cuprates consist of spinons, and the pairing OP has an anisotropic s-wave symmetry, whereas the OP for long-range phase coherence in hole-doped cuprates has the magnetic origin due to spin fluctuations and has the d-wave symmetry.

## V. THE MCS MODEL

In this Section, we describe the MCS model.

Emery and Kivelson pointed out that a finite density of holes in cuprates forms self-organized structures, designed to lower the zero-point kinetic energy.[61] This is accomplished in three stages: a) the formation of charge inhomogeneity (stripes), b) the creation of local spin pairs, and c) the establishment of a phase-coherent state. In general, their approach to the SC in cuprates is correct. However, as we mentioned in Introduction, the phase-coherent state in their model (the stripe model)[11] is established at $T_c$ by the Josephson coupling between charge stripes carrying the spinon SC. In the previous Section, we discussed the coherent OP which has most likely the magnetic origin. This fact is in contradiction with the stripe model.[11] However, the stripe model may fit the data after a serious modification, namely, if instead of the Josephson coupling between stripes we introduce a magnetic interstripe coupling due to spin fluctuations (spin waves).[15] We call this phenomenological model as a MCS (Magnetic Coupling between Stripes) model. In the MCS model, the coherent state of the spinon SC is established due to spin waves into local AF domains, which are excited by *dynamically fluctuating* stripes. Thus, in the MCS model, the SC in cuprates has two mechanisms: along charge stripes and perpendicular to stripes. Charge carriers exhibit different properties in different directions: *fermionic* along charge stripes and *polaronic* perpendicular to stripes. The MCS model is completely based upon experimental facts. Figure 3 shows *schematically* a snapshot of the CuO$_2$ plane in frameworks of the MCS model.

The pseudogap in frameworks of the MCS model is similar to that in the stripe model.[11] By decreasing the temperature, charge stripes are formed at $T^*_{charge} > T_c$, then, at $T^*_{spin} < T^*_{charge}$, AF correlations form AF domains between charge stripes,[7,26,32] and, finally, spin pairs appear along stripes at $T_{pair} < T^*_{spin}$. Thus, in frameworks of the MCS model, above $T_c$, there are three characteristic temperatures: $T^*_{charge} > T^*_{spin} > T_{pair} > T_c$. It is important to note that, between $T_c$ and $T_{pair}$, there are strong fluctuations of the coherent OP (see previous Section), *i.e.* between $T_c$ and $T_{pair}$, some CuO$_2$ planes may be partially SC. The establishment of the long-range phase coherent state at $T_c$ is *mainly* defined by the coherence along *c* axis.

There are two special cases of the SC in cuprates, namely, in NCCO (and its homologues) and YBCO. The SC in the s-wave NCCO cuprate will be considered in Section VII. It is widely believed that the SC in all cuprates occurs into $CuO_2$ planes. The only cuprate which has Cu-O chains is the YBCO family. In YBCO, below $T_c$, chains become SC due to the proximity effect.[62-64] An induced SC gap on chains has to mimic the behavior of the coherent gap. The chain SC occurs probably also due to pairing of spinons since chains are one-dimensional.[11]

The stripe phase in a two-dimensional doped insulator can become either an insulator, as in the nickelates, or a HTSC, as in the cuprates. Typically, quasi one-dimensional metals at low temperatures are insulators due to a CDW order.[11] However, it was shown that the CDW instability is eliminated and the SC is enhanced if the transverse stripe fluctuations have a large enough amplitude.[11] To satisfy this condition, the stripes could oscillate in time or be static and meandering.[11] Thus, the transverse fluctuations of charge stripes are *important* for suppressing a CDW order along the stripes. The stripes in SC cuprates are indeed slowly fluctuating.[19]

## VI. EXPERIMENTAL DATA

In this Section, we present *some* recent experimental data which can be explained by using the MCS model. More experimental data have been discussed in frameworks of the MCS model elsewhere.[16]

Recently, in NMR measurements on near optimally doped YBCO, Gorny *et al.* found that a magnetic field of 14.8 Tesla shifts $T_c$ down by 8 K while the pseudogap above $T_c$ remains unaffected (as measured by $1/(T_1T)$).[65] In frameworks of the MCS models, this means that a magnetic field suppresses the magnetic coupling mechanism between stripes, while the value of the magnetic field of 14.8 Tesla is not enough to destroy the Cooper pairs which exist above $T_c$ on stripes.

In transport measurements in a magnetic field on undoped YBCO ($0.32 \leq x \leq 0.37$), Lavrov *et al.* found that spin fluctuation are playing a major role in the out-of-plane transport regardless of the presence of the Néel order.[66] The Néel temperature *actually* corresponds to the establishment of AF order along the *c* axis.[66] This result is very important for understanding the SC in hole-doped cuprates. By making a parallel between undoped YBCO and SC YBCO, it is reasonable to suggest that spin fluctuations in SC YBCO are also playing a major role in the out-of-plane transport. Consequently, the SC OP for long-range phase coherence in YBCO has most likely the magnetic origin. This result is in a good agreement with the MCS model for hole-doped cuprates.

There is a clear discrepancy among the energy-gap values for different 90 K cuprates, inferred from tunneling measurements.[56] In Bi2212, from Fig. 2(a), there is a distribution of the gap magnitude (21–37 meV). At the same time, the maximum magnitudes of tunneling gaps in Tl2201 (22 meV) and YBCO (21 meV) correspond to the minimum gap magnitude in Bi2212.[56] However, all three cuprates have similar values of $T_c$. In frameworks of the MCS model, by using the phase diagram for hole-doped cuprates we showed that tunneling measurements performed on 90 K cuprates detect two different energy gaps: the pairing gap and the coherent gap, which are identical in conventional superconductors.[56]

In Bi2212, Rener *et. al.* observed by STM at low temperature a large gap into the vortex core, which resembles the pseudogap measured above $T_c$.[67] Figure 4 shows three types of tunneling spectra obtained on under- and overdoped Bi2212 single crystals: at the center of a vortex core at 4.2 K; between vortices at 4.2 K, and above $T_c$ in zero magnetic field. One can see in Fig. 4 that the gaps at the center of the cores are very similar to the pseudogaps measured above $T_c$. In frameworks of the MCS model, inside the vortex cores, the SC with the magnetic pairing is completely suppressed while the spinon SC is almost unaffected, and it is similar to the case above $T_c$. The same effect has been observed on YBCO.[16,68,69] However, in tunneling spectra on YBCO, there is a subgap of 4-6 meV which remains in tunneling spectra obtained at the center of a vortex core.[67,68] We attribute this subgap to a SC gap on chains. The SC on chains *inside* vortex cores is injected along chains from the *outside* of vortex cores.[16]

From femtosecond time-resolved spectroscopy on near optimally doped YBCO, Stevens *et al*. presented an evidence for the existence of two components: band-like and polaronic-like carriers in YBCO.[70] In frameworks of the MCS model, this fact is obvious. On the other hand, none of theoretical models presented in the literature can explain this fact.

In microwave measurements of the complex surface impedance $Z_S = R_S + iX_S$ *in the mixed state* on slightly overdoped Bi2212 single crystals, Hanaguri *et al*. found that $X_S$ which is proportional to the real part of the effective penetration depth increases while there was little changes in $R_S$.[71] The value of the penetration depth is a characteristic of the coherent state while $R_S$ is a characteristic of quasiparticle transport. Thus, in frameworks of the MCS model, the experimental data can be explained by the fact that, in the mixed state, the magnetic mechanism of the SC is seriously damaged while the spinon SC is almost unaffected. The changes in $R_S$ are small since the transport properties of Cooper pairs on stripes are mainly responsible for the value of $R_S$.

At the end of this Section, we turn to the interpretation of INS spectra obtained on LSCO and YBCO. We found that by using the MCS model it is easy to interpret INS spectra. Firstly, it is important to note that an INS measurement is a signal averaging experiment. Secondly, one should notice that, in frameworks of the MCS model, the resonance peak is a magnetic signal of local AF domains in $Cu_2O$ planes and appears in INS spectra at energy approximately equal to $2\Delta_c$ (see Section III), whereas the spin gap occurs along charge stripes and, appears in INS spectra at $E \approx \Delta_p$. Thus, in an INS experiment, the two signals are overlapped. Let us consider different cases depending on hole concentration. (i) In the strongly underdoped regime where $\Delta_p > 2\Delta_c$, the resonance peak occurs inside the spin gap. (ii) In slightly underdoped regime where $\Delta_p \approx 2\Delta_c$, the two signals are superimposed. (iii) At optimal doping and in the overdoped regime where $\Delta_p < 2\Delta_c$, the resonance peak appears above the spin gap. Further, we discuss INS spectra obtained on LSCO and YBCO having different value of $p$. (i) Figure 5 depicts odd and even INS spectra obtained on underdoped YBCO with $T_c$ = 52 K ($p/p_m$ = 0.54) at 5 K.[27] The broad peak at 63 meV in Fig. 5(a) corresponds to the anisotropic s-wave spinon gap, $\Delta_p$, shown in Fig. 2(b), whereas the resonance peak at 25 meV corresponds to spin fluctuations in local AF domains of $Cu_2O$ planes. From the phase diagram shown in Fig. 1, the resonance peak in YBCO with $p/p_m$ = 0.54 should be observed at 25 meV, and the spin gap $\Delta_p$ is equal approximately to 55 meV. Thus, there is a good agreement between the two sets of data. Moreover, in the optical mode shown in Fig. 5(b), the spin gap is observed at 53 meV.[27] (ii) In slightly underdoped LSCO ($x$ = 0.14), INS measurements show only one broad peak at around 20 meV.[72] From Fig. 1, the resonance peak in LSCO with $p/p_m$ = 0.8 should appear at 17.5 meV, and the spin gap $\Delta_p$ is equal approximately to 19 meV. Thus, the two signals are superimposed, and this is in agreement with the INS data.[72] (iii) In slightly overdoped LSCO ($x$ = 0.163), INS measurements show the presence of almost isotropic s-wave spin gap with the magnitude of 6.7 meV and a broad peak at 11 meV.[55] According to the phase diagram shown in Fig. 1, the resonance peak in slightly overdoped LSCO should be observed at 18.2 meV, and the spin gap $\Delta_p$ is equal approximately to 13.4 meV. Unfortunately, the INS data in Ref. 55 are presented only up to 16 meV, and we can not discuss the presence of the resonance peak. However, it seems that the peak at 11 meV corresponds to the anisotropic s-wave $\Delta_p$ shown in Fig. 2(b) with the minimum and maximum magnitudes of 6.7 and 11 meV, respectively (iv) In undoped YBCO$_{6.2}$, the resonance peak is obviously absent but the spin gap is detected in the optical mode at 67 meV,[74] which is in a good agreement with Fig. 1. (v) In very overdoped regime, it is difficult to interpret INS spectra using the phase diagram shown in Fig. 1 since INS spectra become very weak. The shape of the resonance peak shown in Fig. 5(a) is unusually wide. In other INS measurements on underdoped YBCO,[32] the shape of the resonance peak is very narrow, however, the broad peak corresponding to the spin gap is observed at energy which is somewhat larger than it should be according to Fig. 1. In fact, one should take into account that the phase diagram shown in Fig. 1 has to be used carefully, if the value of $p$ is calculated from the $T_c$ value. In addition to this, in YBCO, the situation is more complex than in other cuprates due to the presence of chains.[15,16]

More experimental data have been discussed in frameworks of the MCS model elsewhere.[16] It is possible that there is another model of the HTSC which can explain experimental data better. Unfortunately, we are not aware of such model.

## VII. SUPERCONDUCTIVITY IN S-WAVE NCCO

In this Section, we consider briefly the SC in NCCO and its homologues: $R_{2-x}Ce_xCuO_4$ ($R$ = Pr, Sm, Eu; 0.14 $\leq x$ $\leq$ 0.18). It is widely believed that the NCCO cuprate is electron-doped since it shows a clear *n*-type Hall coefficient through most of the doping level.[75] This family of cuprates presents a very *distinct* set of properties with respect to the other cuprates.[5,76-80] The in-plane resistivity for SC NCCO follows nearly $T^2$ dependence while it has linear dependence in hole-doped cuprates.[80] There is a consensus that the total OP in NCCO has the s-wave symmetry.[5,77-79] This suggests that the SC mediated by spin fluctuations is absent in NCCO.[44] However, the parent compound Nd$_2$CuO$_4$ is the AF insulator with the huge Cu-O-Cu exchange constant within the CuO$_2$ plane.[81] The CuO$_2$ planes are undistorted in NCCO while they are distorted in hole-doped cuprates.[81] Recent INS measurements on NCCO ($x$ = 0.15) show that incommensurate magnetic fluctuations are absent in NCCO.[82,83] Pulsed neutron scattering measurements found no resonance peak in NCCO.[84] The latter facts also demonstrate that the magnetically mediated SC is absent in NCCO. The analysis of many measurements suggests that the SC in NCCO is mediated by phonons.[5,76-80] The change in resistivity transition induced by a magnetic field in NCCO exhibits a different behavior than that in other cuprates.[76] Usually, in hole-doped cuprates, applying a magnetic field causes profound broadening of the resistivity transition (in frameworks of the MCS model, it is clear why, see previous Section), while in NCCO, applying a field ($H \parallel c$-axis) causes the parallel shift of transition curves to lower temperatures.[76] Thus, this picture resembles the case for conventional type-II SCs. At the same time, the upper critical magnetic field, $H_{c2}$, in NCCO is not drastically different than that in other single-layer cuprates.[76] The in-plane coherent length in NCCO is unusually large for cuprates, $\xi \approx 70 - 80$ Å.[5,79] There is an evidence for the existence of two types of charge carriers in NCCO.[85]

Taking into account aforementioned experimental facts we proposed for the SC in NCCO a Phonon Coupling between Stripes (PCS) model.[12,15,16] In the PCS model, stripes carrying spinon SC in order to establish the long-range phase coherence couple to each other due to phonons. After that, new theoretical results show that (i) stripes in electron-doped cuprates are most likely diagonal like that in nickelates; (ii) the charge in electron-doped cuprates is more localized; (iii) stripes in electron-doped cuprates is less mobile in the transverse direction than in hole-doped materials, (iv) the 1/8 effect is not expected to be marked in electron-doped cuprates, and (v) the distance between stripes is larger than in hole-doped cuprates.[86] All these conclusions are in a good agreement with the phonon interstripe mechanism of SC in NCCO.

The absence of SC mediated by spin fluctuations in NCCO is, in fact, surprising since spin waves do propagate into $CuO_2$ planes in undoped NCCO ($x \leq 0.14$) like in other cuprates.[86] However, the increaseof Ce concentration above $x = 0.14$ softens strongly spin waves in NCCO.[83] It seems that magnetic SC mechanism is not used in NCCO because (i) spin waves are too weak in SC NCCO,[83] and/or (ii) stripes in NCCO are most likely diagonal.[86] The presence of two SC mechanisms in frameworks of the PCS model, along and between stripes, implies that, in NCCO, there are also two OPs: a pairing OP and OP for long-range phase coherence, like in hole-doped cuprates (see Section IV). However, both the OPs in NCCO have a s-wave symmetry.[12,15,16,53]

There are strong indications that electron-doped cuprates are, in fact, hole-doped cuprates.[87,88] However, such scenario will not change in general the phenomenological PCS model of SC in s-wave NCCO.

## VIII. ANDERSON'S THEOREM FOR CUPRATES

By analogy with Anderson's theorem for classical SCs, in frameworks of the MCS model, we are able to formulate a similar theorem for cuprates: *any impurity (magnetic or nonmagnetic) doped into $CuO_2$ planes affects both the $T_c$ value and DOS. In hole-doped (electron-doped) cuprates, a nonmagnetic impurity suppresses the SC stronger (milder) than a magnetic impurity.*

In order to show the validity of the theorem let us consider two cases for doped ion: nonmagnetic and magnetic. In the first case, it has been shown that, for example, Zn atoms doped into $CuO_2$ planes create voids ("swiss cheese" model).[89] Consequently, a nonmagnetic ion destroys the local AF order.[89,90] This leads to the destruction of the magnetic interstripe coupling *and* spinon pairs along stripes. Thus, in frameworks of the MCS model, any nonmagnetic ion destroys all elements which lead to the appearance of the HTSC. As a consequence, the value of $T_c$ is lower, and the DOS is seriously affected. One might expect that Zn serves to pin stripes. However, new INS results for a 2% Zn-doped LSCO sample point out that the SC in the cuprate is suppressed *mainly* by "voids" around $Zn^{2+}$ ions.[26]

The effect of a magnetic ion on the SC in hole-doped cuprates is milder than that from a nonmagnetic ion. A magnetic impurity located in AF domains (for example, $Ni^{2+}$) does not destroy completely the local AF order.[90] A magnetic ion which is located on a stripe (for example, $Ni^{3+}$) will affect the spinon pairing along stripes. Probably, the most destructive effect on the HTSC from a magnetic ion is the pinning of fluctuating stripes.[7] So, the effect of magnetic ion on the HTSC must be milder than that of nonmagnetic ion which suppresses completely the SC around it. Indeed, experimentally, it is observed that the partial substitution of nonmagnetic ions $Zn^{2+}$ and $Al^{3+}$ for Cu leads to stronger suppression of the SC than substitution of nominally magnetic Ni ions, which is, obviously, in sharp contrast with the conventional SCs.[90,91] Thus, magnetic and nonmagnetic ions substituted for Cu into $CuO_2$ planes in hole-doped cuprates affect the value of $T_c$ and the DOS but do it differently.

The theorem is valid also for the SC in s-wave NCCO (and its homologues). In frameworks of the PCS model, the effect of magnetic or nonmagnetic ions doped into the $CuO_2$ plane on the SC in NCCO is *opposite* in comparison with the case for hole-doped cuprates. Since, in the PCS model, the value of $T_c$ is defined by the phonon mediated SC, therefore the case of NCCO resembles the case of the conventional SCs: a magnetic ion doped into the $CuO_2$ plane suppresses the SC much stronger than a nonmagnetic ion. Indeed, in optimally doped NCCO, it is experimentally observed that comparing the rates of $T_c$ depression with the substitution of Ni and Zn the rate for the case of Ni is almost ten times the rate found for Zn substitution.[91]

We discuss now the effect on the HTSC from *out-of-plane* impurities in frameworks of the MCS model. Firstly, for out-of-plane substitution, there is no definition, what atom is an impurity in copper-oxides and another one is not. Usually, the bulk value of $T_c$ is considered as the main criterion for it. Thus, the interplanar impurities may affect the value of $T_c$. However, one has to differentiate if the effect originates from pair-breaking or, for example, from structural distortions.[59,92,93] Secondly, for $n$-layer ($n \geq 1$) compounds, there are two different possibilities for out-of-plane doping: (i) between layers and (ii) between $n$-layers (in the case of single-layer cuprates, these two cases coincide). Let us consider briefly the two cases. (i) Interplanar doping. Williams and Tallon showed that, in YBCO, the Ca doping has negligible effect on SC at constant hole concentration.[94] Ca is known to substitute onto the Y site between the two $CuO_2$ layers in YBCO.[94] (ii) Inter-$n$-layer doping. The La substitution for Ba in the insulating BaO layer in YBCO has also negligible effect on SC at constant hole concentration.[94] In another family of cuprates, the change of inter(bi)layer distance by intercalating a long-chain organic compound into Bi2201 and Bi2212 exerts no influence on the $T_c$ value of the cuprates.[95] In Section IV, we described the angle-resolved tunneling measurements on pure and Pb-doped Bi2212 having $T_c$ = 91 K and $T_c$ = 78 K, respectively.[59] The measurements show that, in the Pb-doped Bi2212 cuprate in which Pb ions reside between $CuO_2$ bilayers, Pb affects only the magnitude of the OP for long-range phase coherence, $\Delta_c$, (and, consequently, the $T_c$ value) keeping the magnitude of the pairing OP, $\Delta_p$, unchanged (see Section IV). Since the establishment of long-range phase coherent state at $T_c$ is defined by the coherence along $c$ axis (see Section V), it is possible that, in Pb-doped Bi2212, the *in-plane* value of $T_c$ remains unchanged in comparison with the $T_c$ value in the pure Bi2212, however, the *bulk* value of $T_c$ is affected.[59] In this case, the in-plane DOS is not affected either.[59] Thus, we conclude that if the hole concentration and buckling of $CuO_2$ planes is not changed, the *out-of-plane* substitution does not affect the $\Delta_p$, however, in same cases, it affects the $\Delta_c$ (*i.e.* the *bulk* value of $T_c$).

## XI. DISCUSSION

In this Section, we discuss the most important elements of the MCS model.

Firstly, we begin with the phase diagram for hole-doped cuprates. There is an opinion that the pairing $\Delta_p$ gap shown in Fig. 1 is a normal-state gap. However, tunneling measurements performed by Miyakawa *et al.* show unambiguously that a tunneling gap with the maximum magnitude in under- and overdoped Bi2212 is a SC gap.[52] In *maximum*-gap measurements, the Josephson $I_c R_N$ product (both average and maximum), where $I_c$ is the maximum Josephson current and $R_N$ is the normal resistance of a tunnel junction, *increases* with the decrease of hole concentration while the coherent energy range is also decreasing in the underdoped regime (see Fig. 1). The Josephson strength is a characteristic of the coherent state. The $I_c R_N$ product scales with the $\Delta_p$ and not with the SC $\Delta_c$ gap which scales in its turn with $T_c$. Consequently, the $\Delta_p$ is a SC-like gap. The same conclusion can be made from Andreev-reflection measurements on Bi2223 and Hg2223 performed by Svistunov *et al.*[93] The data show the presence of *two* SC gaps which have different magnitudes and different temperature dependencies.[96]

Do stripes relate directly to the HTSC, or they just co-exist with each other? It is an open question.[31] There is no doubts that stripes are an intrinsic feature in the cuprates (see Section II). But there is no experiment which demonstrates that stripes relate *directly* to the HTSC. From recent measurements on Nd-doped LSCO which show that the stripes are intrinsically metallic,[22] *it seems* that stripes do relate to the occurrence of the SC in the cuprates. Otherwise, the stripes have to be insulating as a consequence of a CDW instability. Recent theoretical investigations show that the pair correlation function of the d-wave channel is suppressed by the stripes, implying the d-wave symmetry of OP along stripes is less favorable than a s-wave symmetry.[97] This result is in a good agreement with the MCS model of the SC in cuprates. Spin fluctuations which exist due to the charge segregation are also intrinsic feature in hole-doped cuprates (see Section III). Any theoretical model of the HTSC has to take into account the presence of charge stripes and spin fluctuations.

There are a few possible explanations for the origin of the magnetic resonance peak observed in *hole*-doped cuprates. On earlier stage, the resonance peak was associated with spin-flip charge carriers excitations across the SC gap.[36] Another possible explanation for the existence of the resonance peak is that it occurs due to electron-hole pairing across the SC gap.[73] It can be a collective spin-wave mode brought about by strong AF correlations.[98] More recently, the resonance peak was associated with the response of the magnetic AF domains to the hole pairing along stripes.[99] In the MCS model, the resonance peak corresponds to spin waves excited in AF domains, which mediate the long-range phase coherence for the spinon SC. There are other explanations for the presence of the magnetic resonance peak in INS spectra obtained on hole-doped cuprates.[36] Some theoretical calculations show that spin fluctuations are too weak to produce the HTSC pairing mechanism. From the discussion in Section III, the coupling strength between charge carriers and spin waves, inferred from INS and infrared spectroscopy data, is sufficient to account for high value of $T_c$.[46] Experimentally, Tsui, Dietz and Walker have been observed in AF $NiO_2$ that electron-magnon interactions at low temperature are much stronger than electron-phonon ones.[100] In frameworks of the MCS model, the pairing mechanism of spinons on stripes is due to pair hopping between the stripe and its AF environment,[11] spin fluctuations in hole-doped cuprates participate only in the establishment of the long-range phase coherence of the spinon SC.

From tunneling measurements, there is a clear evidence for the existence of the d-wave OP in LSCO, YBCO and Bi2212.[101] At the same time, tunneling measurements show that, in NCCO, the d-wave OP is absent,[101] and tunneling spectra resemble a s-wave case.[78] The presence of s-wave component is also found in YBCO[102-106] and Bi2212[107] by tunneling measurements and in LSCO by Andreev-reflection[108] and INS measurements.[55] Torque measurements on Tl2201 show that the total OP consists of the d-wave and s-wave components.[4] Thus, the d-wave and s-wave OPs coexist in hole-doped cuprates.

In single-layer NCCO, the total OP has a s-wave symmetry, and the analysis of many measurements suggests that the SC in NCCO is mediated by phonons.[5,76-80] At the same time, in hole-doped single-layer LSCO, there is an evidence that phonons couple with electrons in the SC state and thus participate *somehow* in the electron-pairing interactions in LSCO.[109] The latter fact suggests that, in LSCO, there is a strong phonon-electron coupling. By trying to unify the MCS and PCS models one can argue that the interstripe coupling mechanism is identical in hole- and electron-doped cuprates and occurs due to interactions between stripes and "unified spin-phonon channel". In electron-doped cuprates, a magnetic contribution to the interstripe coupling is negligible. In most hole-doped cuprates, a phonon contribution to the interstripe coupling is weak. The case of LSCO which has a relatively low value of $T_c$ can be considered as an intermediate one: the phononic contribution is noticeable.[109] The unified model of the SC in the cuprates can be named as a Spin-Phonon Coupling between Stripes (S-PCS) model.

In *strongly* overdoped cuprates, the distance between charge stripes becomes smaller. So, it is reasonable to suggest that, in strongly overdoped regime in hole-doped cuprates, the interstripe coupling may switch from the magnetic channel to the Josephson coupling. This implies that, in strongly overdoped cuprates, the symmetry of total OP in hole-doped cuprates can be *predominantly* a s-wave. It is necessary to carry out experimental and theoretical investigations to verify our proposal.

In frameworks of the MCS and PCS models, in $CuO_2$ planes which are basically identical in all cuprates, there *can* exist three types of SCs: the spinon SC, magnetically mediated SC in hole-doped cuprates and phonon mediated SC in NCCO (and its homologues). Such diversity of different SC mechanisms is a direct consequence of charge inhomogeneity in doped Mott insulators. The OP is not universal in all cuprates. This is an experimental fact. The symmetry of total OP is different for different families of cuprates varying from a pure s-wave in NCCO to the predominant d-wave in Bi2212 and YBCO.[3,5,78,101] By using the MCS model for hole-doped cuprates and PCS model for NCCO (and its homologues) one can explain this difference in the symmetries of total OP in different families of SC cuprates.

## X. CONCLUSIONS

We presented a phenomenological model of the high-$T_c$ superconductivity [a Magnetic Coupling between Stripes (MCS) model] and recent experimental data obtained by different techniques as an experimental evidence in support of the MCS model for hole-doped cuprates.

By analyzing experimental data obtained by different techniques we showed that (i) charge stripes; (ii) spin fluctuations, and (iii) two order parameters are intrinsic futures in hole-doped cuprates. The MCS model for hole-doped cuprates is based upon these facts. In frameworks of the MCS model, the spinon superconductivity occurs above $T_c$ on charge stripes into $CuO_2$ planes. Spinons create pairs on stripes due to pair hopping between the stripe and its antiferromagnetic environment.[11] The long-range coherent state of the spinon superconductivity is established due to spin waves which are excited into local antiferromagnetic domains by dynamically fluctuating charge stripes. Above $T_c$, in frameworks of the MCS model, there are three characteristic temperatures: $T^*_{charge} > T^*_{spin} > T_{pair} > T_c$, which correspond to the formation of charge stripes, antiferromagnetic domains between stripes and spin pairs along stripes, respectively.

We presented also a phenomenological model of superconductivity in s-wave cuprates (NCCO and its homologues) [a Phonon Coupling between Stripes (PCS) model]. In the PCS model for s-wave cuprates, the long-range coherent state of the spinon superconductivity which occurs along charge stripes is established due to phonons. We presented a common theorem for cuprates as an analogy of the Anderson's theorem for conventional superconductors.

We showed that many sets of experimental data can be explained by using the MCS and PCS models. However, it is possible that there is another model of the high-$T_c$ superconductivity which can explain experimental data better. Unfortunately, we are not aware of such model.

## ACKNOWLEDGMENTS

I thank K. Yamada, M. Arai, T. Ekino and R. Deltour for discussions. This work is supported by PAI 4/10.

**FIGURE CAPTIONS**

FIG. 1. Phase diagram for hole-doped cuprates at low temperature: $\Delta_c$ is the magnitude of the OP for long-range phase coherence, and $\Delta_p$ is the magnitude of the OP for pairing.[12,13] The $p_m$ is a hole concentration with the maximum $T_c$. (From Ref. 13).

FIG. 2. (a) Tunneling gap vs. angle in Bi2212 at 5 K: dots (average measured points) and solid line (an assumption of a fourfold symmetry) (from Ref. 58). (b) Shapes of two gaps at low temperature in slightly overdoped Bi2212, shown schematically: the d-wave $\Delta_c$ and anisotropic s-wave $\Delta_p$ (from Refs. 53 & 54).

FIG. 3. Static picture of the $CuO_2$ plane in frameworks of the MCS model (see the text), shown schematically. Spinons form pairs along charge stripes.[11] The SC mechanism in AF insulating stripes is magnetic due to spin fluctuations. The temperature dependencies are shown schematically. The widths of charge and insulating stripes are equal approximately to 8 Å and 16 Å, respectively.[20]

FIG. 4. Tunneling spectroscopy of Bi2212: (a) underdoped and (b) overdoped. The upper and middle spectra in each frame were obtained at 4.2 K and 6 T between vortices and at the center of a vortex core, respectively. The lower spectra in each frame were measured above $T_c$ in zero magnetic field. They illustrate the striking resemblance of the vortex core pseudogap and normal state pseudogap. The spectra have been shifted vertically for clarity. (From Ref. 67).

FIG. 5. Averaged odd (a) and even (b) spin susceptibilities in YBCO at 5 K (from Ref. 27).

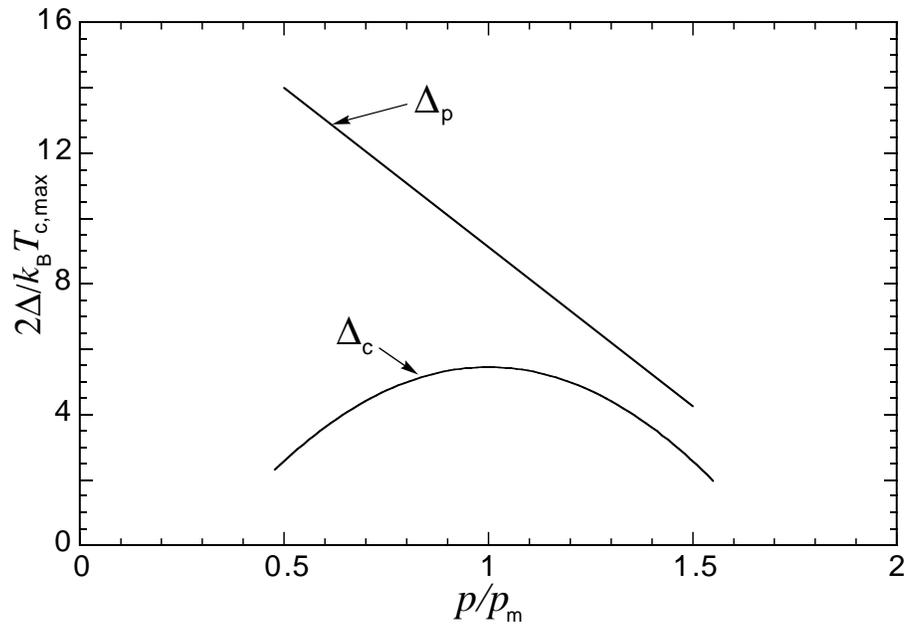

FIG. 1

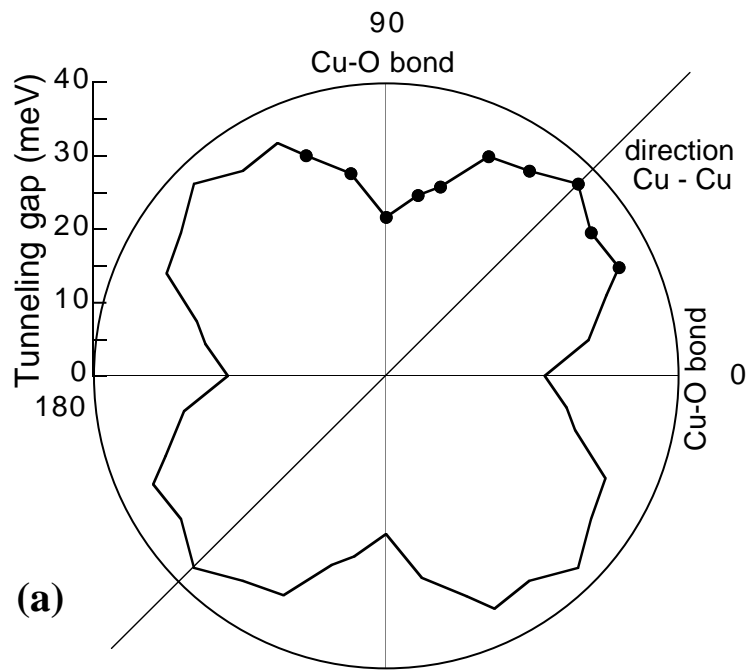
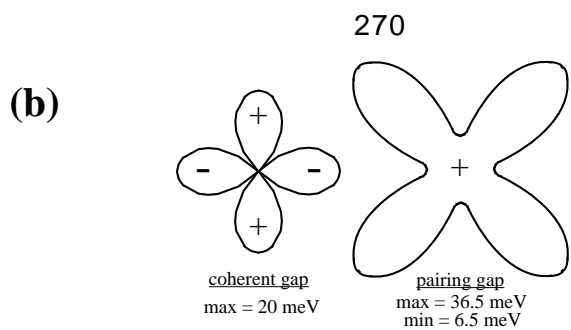

FIG. 2

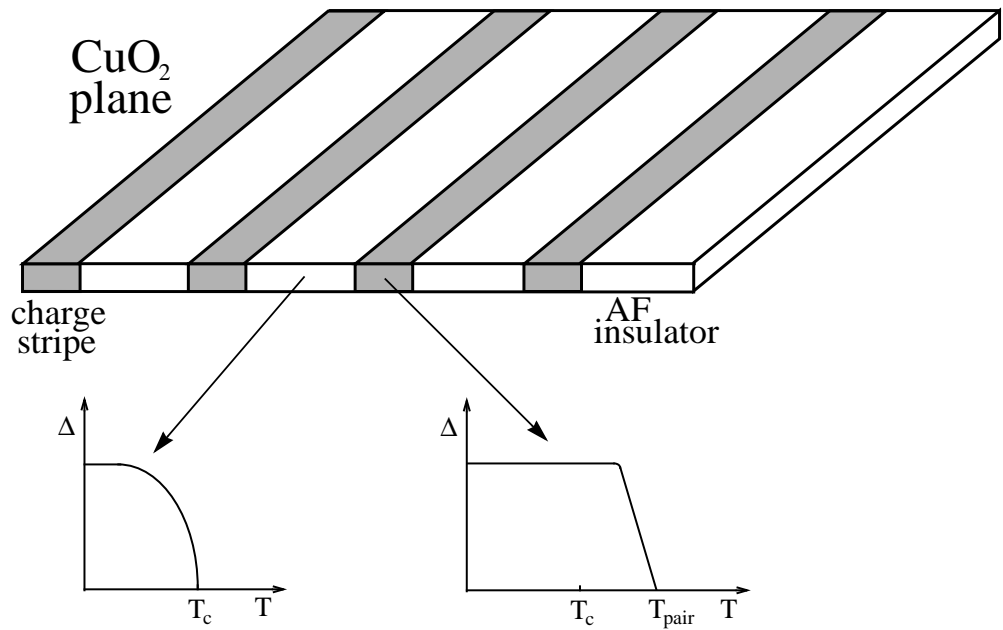

FIG. 3

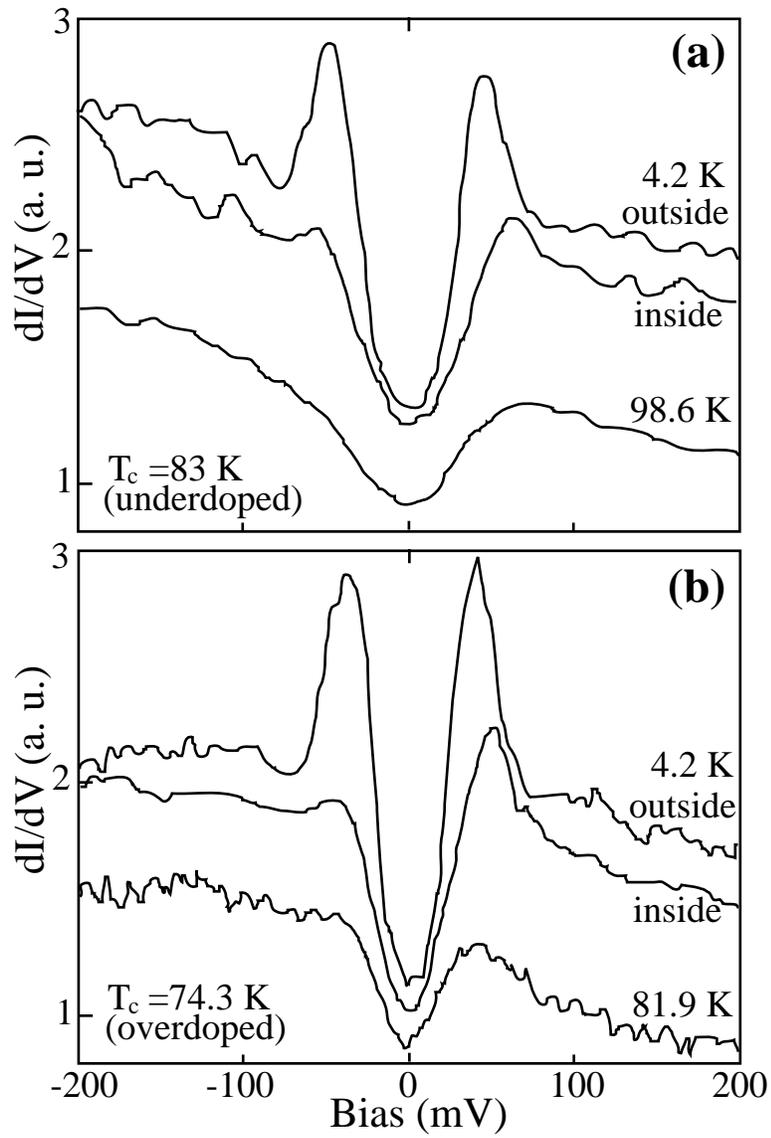

FIG. 4

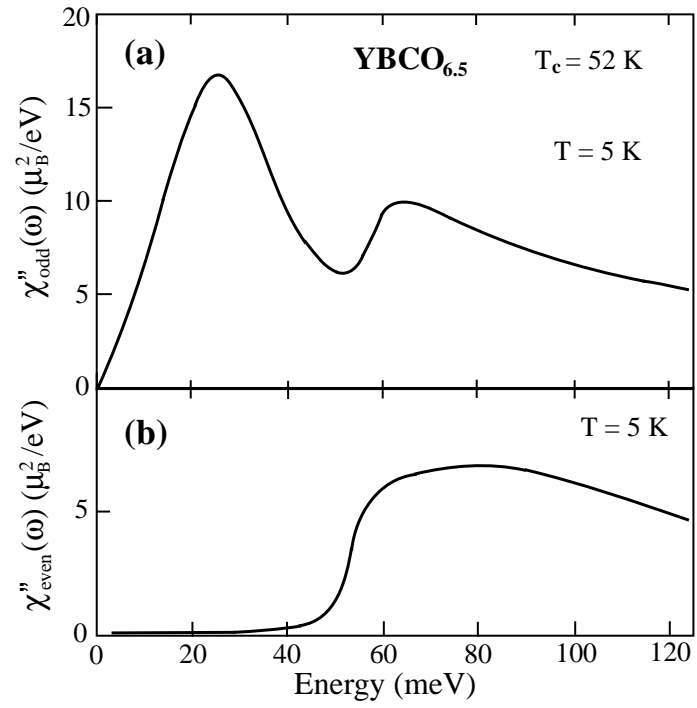

FIG. 5